\documentclass[12pt]{article}
\usepackage{fullpage,
epsfig,psfrag,graphics,amsbsy,amssymb,amsmath,hyperref}
\def\Xint#1{\mathchoice
   {\XXint\displaystyle\textstyle{#1}}%
   {\XXint\textstyle\scriptstyle{#1}}%
   {\XXint\scriptstyle\scriptscriptstyle{#1}}%
   {\XXint\scriptscriptstyle\scriptscriptstyle{#1}}%
   \!\int}
\def\XXint#1#2#3{{\setbox0=\hbox{$#1{#2#3}{\int}$}
     \vcenter{\hbox{$#2#3$}}\kern-.5\wd0}}

\def\dashint{\Xint-}
\usepackage{caption}
\newcommand{\beq}{\begin{equation}}
\newcommand{\eeq}{\end{equation}}
\newcommand{\bea}{\begin{eqnarray}}
\newcommand{\eea}{\end{eqnarray}}

\newcommand{\Tr}{{\rm Tr}}
\newcommand{\tr}{{\rm tr}\ }
\newcommand{\ket}[1]{|#1\rangle}

\def\math{\mathsurround=0pt }
\def\leftrightarrowfill{$\math \mathord\leftarrow \mkern-6mu
 \cleaders\hbox{$\mkern-2mu \mathord- \mkern-2mu$}\hfill
 \mkern-6mu \mathord\rightarrow$}
\def\overleftrightarrow#1{\vbox{\ialign{##\crcr
     \leftrightarrowfill\crcr\noalign{\kern-1pt\nointerlineskip}
     $\hfil\displaystyle{#1}\hfil$\crcr}}}

\let\l=\lambda

 \def\bd{\begin{document}} \def\ed{\end{document}}
\def\ds{\documentstyle} \let\fr=\frac \let\bl=\bigl \let\br=\bigr
\let\Br=\Bigr \let\Bl=\Bigl
\let\bm=\bibitem
\let\na=\nabla
\let\pa=\partial \let\ov=\overline
\def\ft#1#2{{\textstyle{{\scriptstyle #1}\over {\scriptstyle #2}}}}
\def\fft#1#2{{#1 \over #2}}
\def\vp{\varphi}
\def\sst#1{{\scriptscriptstyle #1}}
\def\oneone{\rlap 1\mkern4mu{\rm l}}
\def\td{\tilde}
\def\wtd{\widetilde}
\def\dalemb#1#2{{\vbox{\hrule height .#2pt
        \hbox{\vrule width.#2pt height#1pt \kern#1pt
                \vrule width.#2pt}
        \hrule height.#2pt}}}
\def\square{\mathord{\dalemb{6.8}{7}\hbox{\hskip1pt}}}
\def\wtd{\widetilde}
\def\R{\rlap{\rm I}\mkern3mu{\rm R}}
\def\im{{\rm i}}
\def\tilg{\tilde{g}}
\def\tilF{\tilde{F}}
\def\tilA{\tilde{A}}
\def\varf{\varphi}
\def\tilf{\tilde{\phi}}
\def\tilh{\tilde{h}}
\def\rme{{\rm e}}
\def\ep{\epsilon}
\def\0{{(0)}}
\def\9{{(9)}}
\def\8{{(8)}}
\def\7{{(7)}}
\def\6{{(6)}}
\def\5{{(5)}}
\def\4{{(4)}}
\def\3{{(3)}}
\def\2{{(2)}}
\def\1{{(1)}}
\newcommand{\trace}{{\rm Tr}}
\newcommand{\ub}{\overline{U}}
\newcommand{\vb}{\overline{V}}
\newcommand{\uh}{\widehat{U}}
\newcommand{\vh}{\widehat{V}}
\newcommand{\ubh}{\overline{\widehat{U}}}
\newcommand{\vbh}{\overline{\widehat{V}}}
\newcommand{\lb}{\bar{\l}}
\newcommand{\Fb}{\overline{F}}
\newcommand{\Fh}{\widehat{F}}
\newcommand{\Fbh}{\overline{\widehat{F}}}
\newcommand{\Ab}{\overline{A}}
\newcommand{\Ah}{\widehat{A}}
\newcommand{\Abh}{\overline{\widehat{A}}}
\newcommand{\Gb}{\overline{G}}
\newcommand{\Gh}{\widehat{G}}
\newcommand{\Gbh}{\overline{\widehat{G}}}
\newcommand{\Pb}{\overline{P}}
\newcommand{\Ph}{\widehat{P}}
\newcommand{\Pbh}{\overline{\widehat{P}}}
\newcommand{\Qb}{\overline{Q}}
\newcommand{\Qh}{\widehat{Q}}
\newcommand{\Qbh}{\overline{\widehat{Q}}}
\newcommand{\Bb}{\overline{B}}
\newcommand{\Bh}{\widehat{B}}
\newcommand{\Bbh}{\overline{\widehat{B}}}
\newcommand{\fhns}{\hat{F}^{\rm (NS)}}
\newcommand{\fhrr}{\hat{F}^{\rm (RR)}}
\newcommand{\ahns}{\hat{A}^{\rm (NS)}}
\newcommand{\ahrr}{\hat{A}^{\rm (RR)}}
\newcommand{\hhrr}{\hat{H}^{\rm (RR)}}
\newcommand{\hchi}{\hat{\chi}}
\newcommand{\hphi}{\hat{\phi}}
\newcommand{\htau}{\hat{\tau}}
\newcommand{\cG}{{\cal G}}
\newcommand{\cGb}{\overline{{\cal G}}}
\newcommand{\cH}{{\cal H}}
\newcommand{\cP}{{\cal P}}
\newcommand{\cPb}{\overline{{\cal P}}}
\newcommand{\cQ}{{\cal Q}}
\newcommand{\cQb}{\overline{{\cal Q}}}
\newcommand{\cM}{{\cal M}}
\newcommand{\cN}{{\cal N}}
\newcommand{\cO}{{\cal O}}
\newcommand{\cD}{{\cal D}}
\newcommand{\cL}{{\cal L}}

\newcommand{\vpp}{\mbox{$\langle{\scriptstyle++}\rangle$}}
\newcommand{\vmp}{\mbox{$\langle{\scriptstyle-+}\rangle$}}
\newcommand{\vppp}{\mbox{$\langle{\scriptstyle+++}\rangle$}}
\newcommand{\vmpp}{\mbox{$\langle{\scriptstyle-++}\rangle$}}
\newcommand{\vpmp}{\mbox{$\langle{\scriptstyle+-+}\rangle$}}

\begin{document}
\setlength{\captionmargin}{36pt}
\begin{titlepage}
\begin{flushright}
\phantom{UFIFT-HEP}
\end{flushright}

\vskip 2cm
\begin{center}
\begin{large}
{\bf Color Characters for White Hot String Bits}
\end{large}

\vskip 1cm
{Thomas L. Curtright${}^a$\footnote{E-mail  address: {\tt curtright@miami.edu}}, 
Sourav Raha${}^b$\footnote{E-mail  address: {\tt souravraha@ufl.edu}}, and 
Charles B. Thorn${}^b$\footnote{E-mail  address: {\tt thorn@phys.ufl.edu}} }
\vskip.40cm
${}^a${\it Departments of Physics, University of Miami, Coral Gables, FL 33124 and Washington University, St. Louis MO 63130}
\vskip.20cm
${}^b${\it Institute for Fundamental Theory,\\
Department of Physics, University of Florida,
Gainesville FL 32611}


\vskip 1.0cm
\end{center}

\begin{abstract}\noindent
The state space of a generic string bit model 
is spanned by $N\times N$ matrix creation operators acting on a vacuum state. 
Such creation operators transform in the adjoint representation of the
color group $U(N)$ (or $SU(N)$ if the matrices are traceless). 
We consider a system of $b$ species of bosonic bits and $f$ species
of fermionic bits. The string, emerging in the $N\to\infty$ limit, 
identifies $P^+=mM\sqrt{2}$ with $M$ the bit 
number operator and $P^-=H\sqrt{2}$ with $H$
the system Hamiltonian. We study the thermal
properties of this string bit system in the case $H=0$, which can be considered
the tensionless string limit: the only dynamics is restricting physical
states to color singlets. Then the thermal partition function $\Tr e^{-\beta 
mM}$ can be identified, putting $x=e^{-\beta m}$, with
a generating function $\chi_0^{bf}(x)$, for which the coefficient of $x^n$ 
in its expansion about $x=0$ is the number of
color singlets with bit number $M=n$. This function is a purely group theoretic
object, which is well-studied in the literature.
We show that at $N=\infty$ this system displays a Hagedorn divergence
at $x=1/(b+f)$ with ultimate temperature  $T_H=m/\ln(b+f)$. 
The corresponding function for finite $N$ is perfectly finite
for $0<x<1$, so the $N=\infty$ system exhibits a phase transition
at temperature $T_H$ which is absent for any finite $N$. We demonstrate
that the low temperature phase is unstable above $T_H$. 
The lowest-order $1/N$ asymptotic correction, for $x\to1$ in the
high temperature phase, is computed for large $N$.  Remarkably,
this is related to the number of labeled Eulerian digraphs with
$N$ nodes. Systematic
methods to extend our results to higher orders in $1/N$ are described.
\end{abstract}
\vfill
\end{titlepage}
\section{Introduction}
Recent work on string bits  at finite temperature 
{} \cite{thornhotbits,sourav} explored the nature of a transition from a 
stringy low temperature phase to
a high temperature bit phase. Here we give a follow up
to \cite{sourav}, in which we supplement the numerical analysis of that paper.
and extend consideration to more general string bit models. There
is considerable overlap of many of our mathematical results with earlier work
on gauge theories at zero coupling \cite{sundborg,aharony}.

The state space of string bit models 
\cite{gilest,thornsakh,bergmantsubit,sunthorn,thornspace} is spanned by monomials of matrix 
creation operators $({\bar A_K})_\alpha^{\ \beta}=((A_K)_\beta^\alpha)^\dagger$ acting
on the vacuum $\ket{0}$ annihilated by the $A$'s. In this notation
$\alpha,\beta=1,\ldots, N$ label the fundamental and antifundamental
representations,
respectively, of a color $U(N)$ group. In other words each
string bit lies in the adjoint representation of $U(N)$.
The index $K$
denotes collectively all of the physical properties of a single string bit:
spin, flavor, momentum, etc. A string interpretation can be
given to the color singlet subspace of string bit states,
which can be spanned by states of the form
\bea
\tr({\bar A}_{K_1}\cdots{\bar A}_{K_{k}})\cdots
\tr({\bar A}_{L_1}\cdots{\bar A}_{L_{l}})
\ket{0} \ ,
\eea
where each trace creates a closed chain from the vacuum $\ket{0}$.

For a closed chain to behave like a continuous string (necessary
for Lorentz invariance) the number of
bits in the corresponding trace must be infinite.
But the trace operator description of a multi-string state is unambiguous only
if $N=\infty$ \cite{thornfock}. 
Otherwise by the Cayley-Hamilton theorem, as soon as the number
of bits in a trace is larger than $N$, that trace can be expressed 
in terms of products of smaller traces, such that the total number of
bits remains the same. Of course Nature only requires approximate
Lorentz invariance, in which case $N$ might be finite, albeit
ridiculously large.

The thermal partition function of the string bit system is taken to be
\bea
\Tr e^{-\beta(mM+H)} \ ,\nonumber
\eea 
where $M$ is the bit number operator and $H$ is the Hamiltonian of the 
system. This hypothesis is motivated by the connection to the lightcone
description \cite{goddardgrt} of the emergent
string for which $mM$ is identified as $P^+/\sqrt{2}$ and $H$ is identified
as $P^-/\sqrt{2}$, so that $mM+H$ is identified as $P^0=(P^++P^-)/\sqrt{2}$.
Simple examples of $H$ that produce stringy low energy eigenstates
(see \cite{sunthorn,thornspace}) have the property that non-singlet
eigenstates are separated from singlet eigenstates
by a gap that is infinite
in energy units of the low lying states. Such a low energy phase exhibits the
Hagedorn transition \cite{hagedorn,fubiniveneziano}
at infinite $N$  \cite{thooftlargen,thornult,atickwitten}: 
the partition function diverges
above some finite temperature $T_H=\beta_H^{-1}$. The central result 
revealed by the numerical studies in \cite{sourav} is that this phenomenon is 
present in a much simpler system with $H=0$ but with allowed states restricted
to color singlets  \cite{sundborg}. That is the quantity
\bea
Z&=&\Tr e^{-\beta mM}{\cal P}_{0} \ ,
\eea
where ${\cal P}_0$ projects onto the color singlet subspace,
generically exhibits the Hagedorn phenomenon.  

In \cite{sourav} the focus
was on the simplest string bit model with one bosonic bit 
$a_\alpha^\beta$ and one fermionic bit $b_\alpha^\beta$. 
Then the  bit number 
operator is $M=\tr[{\bar a}a+{\bar b}b]$.
Then $Z$ can be constructed from the character generating function
\bea
\chi(x,\theta)=\Tr e^{-\beta mM+iG_k\theta_k}&=&
\prod_{k,l}\frac{1+xe^{i(\theta_k-\theta_l)}}{1-xe^{i(\theta_k-\theta_l)}} \ ,
\eea 
where $x=e^{-\beta m}$ and $G_k$ is the Cartan subalgebra of $U(N)$.
 To impose singlet restrictions one simply integrates this character over the
group\footnote{More generally the projection
onto a representation whose character is $\chi_R(\theta)$ is achieved by
$Z_R=\int[d\theta]\chi^*_R(\theta)\chi(x,\theta)$, a formula exploited
to good effect in \cite{curtrightz}}
\bea
Z&=&\int [d\theta]\chi(x,\theta)\equiv \chi_0(x) \ ,
\eea
so the coefficient of $x^n$ is precisely the number of color singlet
states with total bit number $M=n$.  In this simplified model the
$Z$ is a purely group theoretic object: the coefficient of $x^n$ counts
the total number of color singlets that reside in the symmetrized
tensor products of adjoints times the antisymmetrized tensor products
of adjoints such that the total number of adjoints is $M=n$.

In this article we consider a general system of string bits
containing $f$ species of fermionic bits and $b$ species of bosonic
bits: 
\bea
\chi^{bf}(x,\theta)=\Tr e^{-\beta mM+iG_k\theta_k}&=&
\prod_{k,l}\frac{(1+xe^{i(\theta_k-\theta_l)})^f}{(1-xe^{i(\theta_k-\theta_l})^b} \ .
\label{bfchar}
\eea 
If the bit creation operators are traceless, and hence are adjoints under
$SU(N)$ rather than $U(N)$, the right side must be multiplied
by $(1-x)^b/(1+x)^f$ which removes the contributions of the traces.

In \cite{sourav} exact results for $\chi_0(x)$ for the case $f=b=1$ were
obtained for $N=2,3,4,5,6$. But for large $N$ a steepest descent
analysis was applied numerically. Fixing $x$, the values of $\theta_k$ which
maximized the integrand were obtained using MATHEMATICA
for many values of $N$ ranging up to 100. Then the logarithm of these
maxima was fit to a form $f_0(x)N^2+f_1(x)N+f_2(x) +f_3(x)\ln x +f_4(x)N\ln N$.
The fit determined $f_0(x)\approx0$ for $x<1/2$ but $f_0(x)>0$ and
monotonically increasing for $x>1/2$. The sharp onset of $f_0(x)$
at $x=1/2$ is the signal of a Hagedorn transition, and the appearance of
an $N^2$ term for $x>1/2$ is a sign of the active role of the adjoint
string bits in the high temperature phase. 
In this sense the transition can be viewed as a deconfinement
transition.

In the present paper we obtain $\chi_0^{bf}(x)$ in the low temperature
regime for $N=\infty$ by exploiting the fact that the uniform distribution
$\theta_k=2\pi k/N$, $k=0,\ldots,N-1$, is a stationary point for all $x$.
For this distribution we can exactly calculate the Hessian double
derivative matrix which determines the Gaussian fluctuations 
about the stationary point, and, because it is a circulant matrix, one
can write down all of it's eigenvalues, and take their large $N$ limit. This
shows that all eigenvalues are negative when $x<1/(f+b)$ 
and $N$ is sufficiently large.
This information allows us to write down the $N\to\infty$ limit
of $\chi_o^{bf}$ for $x<1/(f+b)$:
\bea
\chi_0^{bf}(x)=\prod_{k=1}^\infty\frac{1}{(1-(b+f)x^{2k-1})(1-(b-f)x^{2k})},
\qquad x<\frac{1}{b+f},\quad N=\infty,
\eea
which is a special case of formulas derived long ago in \cite{sundborg,aharony}
in the context of enumerating color singlets in gauge theories at zero
coupling. Although the motivation for these studies was not the study
of string bit models, these authors anticipated most of our results.

For $x>1/(b+f)$ at least one eigenvalue is positive showing that
the uniform distribution is then not a maximum. We have not
succeeded in finding the non-uniform distribution of $\theta$'s
necessary to complete our large $N$ analysis of the high temperature
phase. However, Sundborg's analysis \cite{sundborg} has led to an effective
method \cite{aharony} for obtaining the density of eigenvalues just above the
critical point. In addition, The numerical studies in \cite{sourav}
give a convincing determination of the same distributions.

Our paper is organized as follows.
In Section 2 we define the group characters for the string bit
models studied in this paper and sketch the large $N$ methods
applied to character integrals. In Section 3 we analyze our
string bit models at low temperature. While the main results
of our analysis have been anticipated in \cite{sundborg,aharony},
our methods and motivation are somewhat different, and we think
they offer new insights and clarifications. In Section 4 we
discuss some ideas for solving the high temperature phase.
In Section 5 we develop an effective field theory approach
to generating the $1/N$ expansion in the low temperature phase.
This perturbation expansion breaks down near the critical point,
but we offer a partial resummation procedure generated by the Dyson
equation, which at least makes the propagator finite at the
critical point. In Section 6, we analyze the thermal partition
function near infinite temperature $x\approx1$ and discuss
its relation to the problem of counting Eulerian digraphs.
We close with a concluding Section 7.

\section{$SU(N)$ Characters and Large $N$}
In a general string bit model the bits are 
adjoint creation operators, $b$ bosonic and $f$ fermionic.
Rearranging the products on the right of (\ref{bfchar}),
the $U(N)$ character of this model can be written
\bea
\chi^{bf}(x,\theta_1,\ldots,\theta_N)&=&\frac{(1+x)^{Nf}}{(1-x)^{Nb}}
\prod_{k<l}\frac{(1+xe^{i\theta_{kl}})^f(1+xe^{-i\theta_{kl}})^f}
{(1-xe^{i\theta_{kl}})^b(1-xe^{-i\theta_{kl}})^b} \ ,
\label{subitchar}
\eea
where $\theta_{kl}\equiv\theta_k-\theta_l$.  Here the coefficient
of $x^n$ is the character of all states with bit number $M=n$.
The $\theta$'s parameterize the transformations due to the
Cartan sub-algebra.
String bit models rely on color
confinement to yield string models. This confinement is a salient
feature of the stable string bit models \cite{sunthorn}, 
in which it arises from the dynamics. For example, in the case $b=f=1$,
this is accomplished by the Hamiltonian
\bea
H&=&\frac{T_0}{2mN}\tr\left[({\bar a}^2 -i{\bar b}^2)a^2
-({\bar b}^2-i{\bar a}^2)b^2+({\bar a} {\bar b}
+{\bar b} {\bar a})ba+({\bar a} {\bar b}
-{\bar b} {\bar a})ab\right] \ ,
\label{susybitham}
\eea
where $T_0$ is the rest tension of the emergent string.
In this paper we replace this dynamics 
with the restriction of allowed states to color singlets,
setting $H=0$. Since $H$ is proportional to $T_0$, we can say
that $H=0$ describes the tensionless limit of the emergent
string theory. 
The projection onto color singlet states is achieved
by averaging the
character (\ref{subitchar})  over the group.
\bea
\chi_0^{bf}&=&D^{-1}\int [d\theta] 
\frac{(1+x)^{Nf}}{(1-x)^{Nb}}
\prod_{k<l}\frac{(1+xe^{i\theta_{kl}})^f(1+xe^{-i\theta_{kl}})^f}
{(1-xe^{i\theta_{kl}})^b(1-xe^{-i\theta_{kl}})^b}\\
D&=&\int [d\theta] \equiv \int d\theta_1\cdots d\theta_N
\prod_{k<l}{(1-e^{i\theta_{kl}})(1-e^{-i\theta_{kl}})} \ .
\label{singletchar}
\eea
\subsection{Large $N$ Approximation}
Notice that $\ln\chi$, the logarithm of the integrand for $\chi_0$, 
can be expressed as a double sum, each ranging from
$1$ to $N$: it generically should be of order $N^2$ as $N\to\infty$. This
justifies approximating the integral by a steepest descent
evaluation valid for large $N$. Define
\bea
L(\theta)\equiv \sum_{k>l}\ln\frac{(1-e^{i\theta_{kl}})(1-e^{-i\theta_{kl}})(1+xe^{i\theta_{kl}})^f(1+xe^{-i\theta_{kl}})^f}
{(1-xe^{i\theta_{kl}})^b(1-xe^{-i\theta_{kl}})^b} \ ,
\eea
where $f$ is the number of fermion bits,
$b$ the number of bosonic bits, and $\theta_{kl}\equiv \theta_k-\theta_l$.
We seek a maximum for $L$ which requires it to be stationary:
\bea
0&=&\frac{\partial L}{\partial\theta_k}
=\sum_{l\neq k}\left[\cot\frac{\theta_{kl}}{2}
-\frac{2(b+f)x(1+x^2)\sin\theta_{kl}+2(b-f)x^2\sin2\theta_{kl}}{1+x^4-2x^2\cos2\theta_{kl}}
\right] \ .
\label{stationarity}
\eea
For large $N$ one should be able to treat the sum over $l$ as an integral
(see for example \cite{brezin,grosswitten,wadia}),
by first ordering the $\theta$'s so that $\theta_1<\theta_2<\cdots<\theta_N$
and then introducing a density function $\rho$ via 
\bea
\sum_lS_l\to N\int_{-\pi}^\pi d\theta\rho(\theta)S(\theta) \ , \qquad \int d\theta\rho(\theta)=1 \ ,
\eea
after which the stationarity condition becomes an integral equation
\bea
0={\dashint} d\theta^\prime\rho(\theta^\prime)\left[\cot\frac{\theta-\theta^\prime}{2}
-\frac{2(f+b)x(1+x^2)\sin(\theta-\theta^\prime)+2(b-f)x^2\sin2(\theta-\theta^\prime)}{1+x^4-2x^2\cos2(\theta-\theta^\prime)}\right] \ ,
\label{inteq}
\eea
where $\dashint$ denotes the principal value prescription for $\theta^\prime=\theta$.
Correspondingly
\bea
L&\to&L_0\equiv N^2\int_{\theta>\theta^\prime}d\theta\rho(\theta)d\theta^\prime\rho(\theta^\prime)\ln\frac{|1-e^{i(\theta-\theta^\prime)}|^2|1+xe^{i(\theta-\theta^\prime)}|^{2f}}{|1-xe^{(\theta-\theta^\prime)i}|^{2b}} \ .
\eea
The uniform distribution $\rho=1/(2\pi)$, for $-\pi<\theta<\pi$ solves 
(\ref{inteq}) for all $x$. However this stationary point may not
be the global maximum. Indeed for this solution, it is easy to show
that $L=0$, and any solution for which $L_0>0$ would dominate it.

In fact $\theta_k=2\pi k/N$, $k=1,\ldots,N$ is a solution of 
(\ref{stationarity}) before passing to the continuum limit.
To see this, call the summand in the equation evaluated on these 
$\theta$'s $S(k-l)$:
\bea
S(k-l)&=&\cot\frac{2\pi (k-l)}{2N}\nonumber\\
&&-\frac{2(f+b)x(1+x^2)\sin(2\pi(k-l)/N)+2(b-f)x^2\sin(4\pi(k-l)/N)}{1+x^4-2x^2\cos(4\pi(k-l)/N)} \ .
\eea
Then by inspection we have $S(n+N)=S(n)$ and $S(n)=-S(-n)$. Then
\bea
\sum_{l\neq k}S(k-l)&=&\sum_{l=1}^{k-1}S(k-l)+\sum_{l=k+1}^NS(k-l)
=\sum_{l=1-k}^{-1}S(-l)+\sum_{l=1}^{N-k}S(-l)\nonumber\\
&=&\sum_{l=1}^{N-1}S(-l)=\sum_{l=1}^{N-1}S(-(N-l))=\sum_{l=1}^{N-1}S(l)
=-\sum_{l=1}^{N-1}S(-l) \ .
\eea
From the second line we see that $\sum S(-l)=-\sum S(-l)$ which implies
$\sum S(-l)=0$. This establishes that $\theta_k=2\pi k/N$ solves
the stationarity equations. 
\section{Low temperature solution at $N=\infty$}
Actually, one can straightforwardly evaluate the integrand for $\theta_k
=2\pi k/N$, with $k=1,2,\ldots,N$, i.e. for the uniform distribution of
the $\theta$'s, without passing to the continuum approximation of the sums.
\bea
L(\theta)&\to& \frac{1}{2}\sum_{k\neq l}\left[\ln|1-e^{2\pi i(k-l)/N}|^2
+\ln|1+xe^{2\pi i(k-l)/N}|^{2f}-\ln|1-xe^{2\pi i(k-l)/N}|^{2b}
\right]\nonumber\\
&=&\frac{N}{2}\sum_{k=1}^{N-1}\left[\ln|1-e^{2\pi ik/N}|^2
+\ln|1+xe^{2\pi ik/N}|^{2f}-\ln|1-xe^{2\pi ik/N}|^{2b}
\right] \ .
\eea
Each of the sums can be evaluated as follows:
\bea
\sum_{k=1}^{N-1}\ln(1-ue^{2\pi ik/N})
=-\sum_{n=1}^\infty\frac{u^n}{n}\sum_{k=1}^{N-1}e^{2\pi i nk/N} \ .
\eea
The inner sum is $N-1$ for $n$ an integer multiple of $N$ and if not
it is $-1$. Then we have
\bea
\sum_{k=1}^{N-1}\ln(1-ue^{2\pi ik/N})=-N\sum_{l=1}^\infty\frac{u^{lN}}{lN}
+\sum_{n=1}^\infty \frac{u^n}{n}=\ln\frac{1-u^N}{1-u} \ .\eea
And we need this formula for $u=1$ when it gives $\ln N$, and for 
$u=-x$ and $u=x$. Putting these things together gives
\bea
L&\to&N\ln N+Nf\ln\frac{1-(-)^Nx^N}{1+x}-Nb\ln\frac{1-x^N}{1-x}\nonumber\\
&=&N\ln N+N\ln\frac{(1-x)^b}{(1+x)^f}+O(x^N) \ .
\eea
The $N\ln N$ is just the $x=0$ limit representing the maximum of the
group measure factor which is dropped upon normalizing the character.
The integral we are approximating is multiplied by
a factor $(1+x)^{(N-1)f}/(1-x)^{(N-1)b}$ for $SU(N)$ or with the -1 removed
for $U(N)$. In the first case the $N$ in the second term on the right
will be replaced by 1, and in the second case it will be completely
cancelled. Thus the net contribution of the maximum of the integrand to the
character will be a factor $(1-x)^b/(1+x)^f$ for $SU(N)$ and 1 for $U(N)$.
The rest of the large $N$ approximation to the character
comes from the Gaussian integral
of the fluctuations about the uniform distribution, to which we now turn.
\subsection{Gaussian fluctuations}
One can form the second derivative matrix of $L$ (exhibited here only for $f=b=1$):
\bea
\frac{\partial^2L}{\partial\theta_m\partial\theta_k}
&=&\sum_{l\neq k}(\delta_{km}-\delta_{lm})\bigg[-\frac{1}{2}
\csc^2\frac{\theta_k-\theta_l}{2}
-\frac{4x(1+x^2)\cos(\theta_k-\theta_l)}{1+x^4-2x^2\cos2(\theta_k-\theta_l)}
\nonumber\\
&&+\frac{16x^3(1+x^2)\sin(\theta_k-\theta_l)\sin2(\theta_k-\theta_l)}{(1+x^4-2x^2\cos2(\theta_k-\theta_l))^2}\bigg]\nonumber\\
\frac{\partial^2L}{\partial\theta_k\partial\theta_k}&=&-\sum_{l\neq k}\bigg[\frac{1}{2}
\csc^2\frac{\theta_k-\theta_l}{2}
+\frac{4x(1+x^2)\cos(\theta_k-\theta_l)}{1+x^4-2x^2\cos2(\theta_k-\theta_l)}
\nonumber\\
&&-\frac{16x^3(1+x^2)\sin(\theta_k-\theta_l)\sin2(\theta_k-\theta_l)}{(1+x^4-2x^2\cos2(\theta_k-\theta_l))^2}\bigg],\qquad m=k\\
\frac{\partial^2L}{\partial\theta_m\partial\theta_k}&=&
\frac{1}{2}
\csc^2\frac{\theta_k-\theta_m}{2}
+\frac{4x(1+x^2)\cos(\theta_k-\theta_m)}{1+x^4-2x^2\cos2(\theta_k-\theta_m)}
\nonumber\\
&&-\frac{16x^3(1+x^2)\sin(\theta_k-\theta_m)\sin2(\theta_k-\theta_m)}{(1+x^4-2x^2\cos2(\theta_k-\theta_m))^2},\qquad m\neq k \ .
\eea
To test the stability of the uniform $\theta$ distribution, one can simply
substitute $\theta_n\to 2\pi n/N$ in these formulas which makes the
second derivative matrix a circulant matrix for which the eigenvalues 
can be written down:
\bea
K_{kk}\equiv c_0&=&-\sum_{l=1}^{N-1}\bigg[\frac{1}{2}
\csc^2\frac{\pi l}{N}
+\frac{4x(1+x^2)\cos(2\pi l/N)}{1+x^4-2x^2\cos(4\pi l/N)}
\nonumber\\
&&-\frac{16x^3(1+x^2)\sin(2\pi l/N)\sin(4\pi l/N)}{(1+x^4-2x^2\cos(4\pi l/N))^2}\bigg],\qquad m=k\nonumber\\
K_{km}\equiv c_{k-m}&=&
\frac{1}{2}
\csc^2\frac{\pi(k-m)}{N}
+\frac{4x(1+x^2)\cos(2\pi(k-m)/N)}{1+x^4-2x^2\cos(4\pi(k-m)/N)}
\nonumber\\
&&-\frac{16x^3(1+x^2)\sin(2\pi(k-m)/N)\sin(4\pi(k-m)/N)}{(1+x^4-2x^2\cos(4\pi(k-m)/N)))^2} \ ,\quad m\neq k \ .
\eea
Then the eigenvalues $\mu_n$ of the matrix $K$ are given by
\bea
\mu_n&=&\sum_{l=0}^{N-1}c_le^{-2\pi i nl/N}=
\sum_{l=0}^{N-1}c_l\cos\frac{2\pi  nl}{N}
=-\sum_{l=1}^{N-1}c_l\left(1-\cos\frac{2\pi nl}{N}\right),
\label{2point}\eea 
for which the eigenvectors are
\bea
V^n_k&=&\frac{e^{2\pi ink/N}}{\sqrt{N}}
,\qquad V^{m\dagger}V^n=\frac{1}{N}\sum_{k=1}^N
e^{2\pi i(n-m)k/N}=\delta_{mn} \ .
\eea
In the large $N$ limit the sum in (\ref{2point}) goes over into an integral.
Put $\theta=2\pi l/N$ and regard $c_l$ as a function $c(\theta)$. Then
\bea
\mu_n&\to&-\frac{N}{2\pi}\int_0^{2\pi}d\theta c(\theta)(1-\cos n\theta) \ .
\eea
This integral can be done by changing variables to $z=e^{i\theta}$ whence
the integral becomes a closed contour integral which can be evaluated
by residues. But first we can simplify the job by recognizing that
by construction (returning to general $b,f$)
\bea
c(\theta)=-\frac{d}{d\theta}\left[\cot\frac{\theta}{2}
-\frac{2(f+b)x(1+x^2)\sin(\theta)+2(b-f)x^2\sin2\theta}{1+x^4-2x^2\cos2\theta}\right] \ . 
\eea
Inserting this into the formula for $\mu_n$ and integrating by parts gives
\bea
\frac{\mu_n}{N}&=&-\frac{n}{2\pi}\int d\theta\sin n\theta
\bigg[\cot\frac{\theta}{2}-\frac{(f+b)+(b-f)\cos(\theta-i\ln x)}{2\sin(\theta-i\ln x)}\nonumber\\
&&-\frac{f+b+(b-f)\cos(\theta+i\ln x)}{2\sin(\theta+i\ln x)}\bigg] \ .
\eea
Now changing variables to $z=e^{i\theta}$ leads to the contour integral
\bea
-\frac{\mu_n}{N}&\to&\frac{n}{2\pi i}\oint 
\frac{dz}{z^{n+1}}\frac{z^{2n}-1}{2}\bigg[
\frac{z+1}{z-1}-\frac{(f+b)xz}{x^2z^2-1}+\frac{(f-b)x^2z^2}{x^2z^2-1}
\nonumber\\
&&-\frac{(f+b)xz}{z^2-x^2}
+\frac{(f-b)x^2}{z^2-x^2}\bigg] \ .
\eea
The contour starts out on the unit circle, so the only contributing poles
are the $(n+1)$th order one at $z=0$ and the last two terms in square brackets
with the simple poles at $z=\pm x$. The former contributes an amount
$n(1-(x^{n}+x^{-n})(f+b)/2)$ when $n$ is odd, and $n(1-(b-f)(x^n+x^{-n})/2)$ when $n$ is even. The latter produces  $-n(x^{n}-x^{-n})(f+b)/2$ for $n$ odd
and $-n(x^{n}-x^{-n})(b-f)/2$ for $n$ even.
Thus
\bea
-\frac{\mu_n}{N}&\to& \begin{cases} n(1-(b+f)x^{n})& n\quad {\rm odd}\\
n(1-(b-f)x^n) & n\quad {\rm even}\end{cases} \ ,\qquad N\to\infty \ .
\label{quadratic}
\eea
These eigenvalues are actually doubled because the limit $N\to\infty$ can
be taken either with $n$ fixed or $N-n$ fixed with identical results.
The saddle point evaluation of the character integral about the
uniform $\theta$ distribution includes a factor of $(\det^\prime K)^{-1/2}$,
where the prime indicates the deletion of the zero eigenvalue,
multiplying the maximum value of the integrand. Of course
$\det^\prime K \propto \prod_{n=1}^\infty[\mu_n^2/(nN)]$ where we drop
multiplicative constants since the character is normalized to
unity at $x=0$. Then the $U(N)$ superstring bit character in the limit
$N\to\infty$ is
\bea
Z_{N=\infty}&=&\prod_{k=1}^\infty\frac{1}{(1-(b+f)x^{2k-1})(1-(b-f)x^{2k})} \ ,
\eea
valid whenever the uniform distribution maximizes the character
integrand, that is for $x<1/(f+b)$. The $SU(N)$ character is obtained by multiplying
this by $(1-x)^b/(1+x)^f$.

Interesting special cases of this formula are the simplest superstring bit model $f=b=1$
where the answer is $\prod(1-2x^{2k-1})^{-1}$; the single bosonic bit
$b=1,f=0$ with answer  $\prod(1-x^{k})^{-1}$; and the single fermionic bit
$f=1,b=0$:
\bea
\prod_{k=1}^\infty\frac{1}{(1-x^{2k-1})(1+x^{2k})}
&=&\prod_{k=1}^\infty\frac{1-x^{2k}}{(1-x^{2k-1})(1-x^{4k})}
=\prod_{k=1}^\infty\frac{1-x^{2(2k-1)}}{1-x^{2k-1}}\nonumber\\
&=&\prod_{k=1}^\infty({1+x^{2k-1}}),\qquad f=1,\quad b=0 \ .\eea
For the case $b=f=1$ we used our formula to generate the singlet
multiplicities of levels up to $M=60$. The results agree exactly 
with the selected multiplicities obtained in \cite{gaolisun}, by counting
states in the multitrace basis. At $M=60$ this agreement is with
all 18 digits of the integer $837224873334502342$. To get the
asymptotic multiplicity for very large bit number, our formula can be approximated by its behavior near $x=1/(b+f)$:
\bea
Z_{N=\infty}\approx\frac{1}{1-(b+f)x}\prod_{k=1}^\infty\frac{1}{(1-(b+f)^{-2k})
(1-(b-f)(b+f)^{-2k})} \ .
\eea
The asymptotic multiplicity is the coefficient of $x^M$ in this approximate formula:
\bea
g(M)&\sim& (b+f)^M\prod_{k=1}^\infty\frac{1}{(1-(b+f)^{-2k})
(1-(b-f)(b+f)^{-2k})}\\
&\to& 2^M\prod_{k=1}^\infty\frac{1}{1-2^{-2k}}\approx 1.452353642449597\cdot2^M
\ ,\qquad b=f=1.
\eea
also in agreement with \cite{gaolisun}.

In order to assess the situation at finite $N$, we employed MATLAB to 
numerically compute the eigenvalues. Such a study, in the case $b=f=1$ 
shows that, for $x<1/2$ and large
enough $N$, except for a single zero eigenvalue due to translational
invariance, all the eigenvalues are negative. For lower values of $N$,
positive eigenvalues develop even for $x<1/2$. Indeed when $x$ is close
to $1/2$, $N$ must be taken quite large to assure only negative
eigenvalues: for instance, for $x=0.499$, $N$ must be greater than $501$
to ensure that all eigenvalues are negative. 
The negativity of all eigenvalues shows that the uniform
solution is then a MAXIMUM (at least locally). 
For $x>1/2$ the same numerical study shows that at
least one eigenvalue becomes positive, indicating that the uniform distribution
is a saddle point and hence not a true maximum.
  
Similarly, it is straightforward to numerically search 
for the global maximum of
$L(\theta)$ and fit the results to a model $N$ dependence
$f_0(x)N^2+f_1(x)N\ln N+f_2(x)N+f_3(x)\ln N+f_4(x)$ \cite{sourav}. Such an analysis supports
the conclusion that
the uniform distribution is a global maximum ($f_0(x)=0$) for $x<1/2$.
For $x>1/2$ positive values for $f_0(x)$ were obtained, with a
nonuniform $\theta$ distribution $\rho(\theta)$, which vanishes outside
an interval $-a(x)<\theta<a(x)$, where $a(x)$ decreases as $x$
approaches 1. Our efforts to find this nonuniform distribution 
analytically have so far come up short, but see the discussion
of this issue in \cite{sundborg,aharony}.
\section{Toward a high temperature solution as $N\to\infty$}
The method that successfully solved the unitary one matrix problem
\cite{grosswitten,wadia} was couched in terms of complex analysis. 
Following this idea for our problem, we define a
function $F(z)$ of the complex variable $z$ by (we assume
supersymmetry ($f=b$) for simplicity)
\bea
F(z)&=&{\int_{-\alpha}^\alpha}d\theta^\prime\rho(\theta^\prime)\left[\cot\frac{z-\theta^\prime}{2}
-\frac{4bx(1+x^2)\sin(z-\theta^\prime)}{1+x^4-2x^2\cos2(z-\theta^\prime)}\right]
\ .\eea
Clearly this function is periodic $F(z+2\pi)=F(z)$ and has branch points,
joined by a finite cut on the real axis, 
at $z=\pm\alpha$ mod $2\pi$ due to the first
term in square brackets. In addition the second term causes branch points
at $z=\pm\alpha+i\ln x$ mod $\pi$ and at $z=\pm\alpha-i\ln x$ mod $\pi$.
These additional branch points are joined by finite cuts which overlap
when $\alpha>\pi/2$ but do not overlap when $\alpha<\pi/2$. The normalization
$\int d\theta\rho(\theta)=1$ implies that $F(z)\to \mp i$ when
$z\to \pm i\infty$.

If such an analytic function can be found, with the property
that $F(\eta\pm i\epsilon)=\mp2\pi i\rho(\eta)$ for real $\eta$
in the range  $-\alpha<\eta<\alpha$, then $\rho(\eta)$ will solve the
integral equation. In the case of the unitary matrix problem a fairly simple
guess involving the square root function led to the unique answer.
Here the guesswork is more challenging.

For a start, notice that the second term in square brackets may be expressed in
partial fractions as
\bea
&&\hskip-.4in\frac{4x(1+x^2)\sin(z-\beta)}{1+x^4-2x^2\cos2(z-\beta)}\nonumber\\
&=&\frac{1}{2}\bigg[\cot\frac{z-\beta-i\ln x}{2}-\cot\frac{z-\beta-\pi-i\ln x}{2}\nonumber\\
&&\hskip1in+\cot\frac{z-\beta+i\ln x}{2}-\cot\frac{z-\beta-\pi+i\ln x}{2}
\bigg]\nonumber\\
&=&\frac{1}{\sin(z-\beta-i\ln x)}+\frac{1}{\sin(z-\beta+i\ln x)} \ .
\eea
The next to last line shows that we can express the function $F(z)$
in terms of a simpler function 
\bea
g(z)&\equiv&\int_{-\alpha}^{\alpha}d\theta^\prime\rho(\theta^\prime)\cot\frac{z-\theta^\prime}{2}
\eea
via
\bea
F(z)&=&g(z)-\frac{b}{2}\left[g(z-i\ln x)+g(z+i\ln x)\right.\nonumber\\
&&\qquad \left. -g(z-\pi-i\ln x)-g(z-\pi+i\ln x)\right],\qquad f=b.
\eea
It remains to be seen whether this aids the problem of guessing the answer!
\section{Effective Field Theory: $1/N$ Expansion}
As we have seen, for $N$ large enough,
the uniform distribution $\theta^0_k={2\pi k/N}$ is a maximum
of $L(\theta)$ for $x<1/(b+f)$. We can then write 
$\theta_k=\theta^0_k+\delta\theta_k$ and 
expand $L$ in a series of powers of $\delta\theta$:
\bea
L(\theta)&=&L(\theta^0)+\frac{1}{2}\delta\theta_k\delta\theta_lK_{kl}
+\frac{1}{3!}\delta\theta_k\delta\theta_l\delta\theta_m{\cal V}^3_{klm}
+\cdots,
\label{fluct}
\eea 
where $K_{kl}=c_{k-l}$, the second derivative matrix has already been
introduced in Eq.(\ref{2point}). In the discussion following that equation
we noted that the eigenvalues of $K$ in the large $N$ limit are proportional to 
$N$ and nonpositive for $x<1/(b+f)$. Truncating the expansion at
order $\delta\theta^2$ and integrating over the fluctuations is
the approximation used to obtain the large $N$ limit of the partition
function.

The cubic and higher terms involve higher derivatives of $L$
\bea
{\cal V}^n_{k_1k_2\cdots k_n}&=&\frac{\partial^nL}{\partial\theta_{k_1}
\cdots\partial\theta_{k_n}}\bigg|_{\theta=\theta^0} \ ,
\eea
again evaluated at the uniform $\theta$ distribution. The $\delta\theta$'s 
can be expanded in eigenvectors of $K_{kl}$, 
\bea
\delta\theta_k&=&\sum_{n=0}^{N-1}\lambda_nV^n_k=
\frac{1}{\sqrt{N}}\sum_{n=0}^{N-1}\lambda_n e^{2\pi ink/N},
\eea
and the expansion (\ref{fluct}) recast as a series in powers of $\lambda$.
Then the coefficient of $\lambda^m/m!$ is
\bea
{\tilde{\cal V}}^m_{n_1\cdots n_m}&=&\frac{1}{N^{m/2}}\sum_{k_1\cdots k_m}
e^{2\pi i(n_1k_1+\cdots+n_mk_m)/N}{\cal V}^m_{k_1\cdots k_m} \ .
\eea
For $N\to\infty$, one can show that ${\tilde{\cal V}}^m$ scales as
$N^{2-m/2}$. We have already seen that the eigenvalues of $K$ scale
as $N$, which is this scaling law for $m=2$. 
To illustrate how this scaling law works at higher order, we work
it out explicitly for $m=3$. Start by noticing that 
$L=(1/2)\sum_{k\neq l}{\cal L}(\theta_k-\theta_l)$. Then
\bea
\frac{\partial^3L}{\partial\theta_{k_1}\partial\theta_{k_2}\partial\theta_{k_3}}
&=&\frac{1}{2}\sum_{k\neq l}\prod_{j=1}^3(\delta_{k_jk}-\delta_{k_jl})
{\cal L}^{\prime\prime\prime}(\theta_k-\theta_l) \ ,
\eea
where the primes denote derivative with respect to the argument.
Next substitute $\theta_k\to 2\pi k/N$ and calculate
\bea
{\tilde{\cal V}}^3_{n_1n_2n_3}&=&\frac{1}{2N^{3/2}}\sum_{k\neq l}\prod_{j=1}^3
(e^{2\pi i n_jk/N}-e^{2\pi i n_jl/N}){\cal L}^{\prime\prime\prime}\left(\frac{2\pi(k-l)}{N}\right)\nonumber\\
&=&\frac{1}{2N^{3/2}}\sum_{l=1}^N\sum_{k=1}^{N-1}\prod_{j=1}^3
(e^{2\pi i n_j(k+l)/N}-e^{2\pi i n_jl/N}){\cal L}^{\prime\prime\prime}\left(\frac{2\pi k}{N}\right)\nonumber\\
&=&\frac{1}{2N^{3/2}}\sum_{l=1}^Ne^{2\pi il(n_1+n_2+n_3)/N}\sum_{k=1}^{N-1}\prod_{j=1}^3
(e^{2\pi i n_j k/N}-1){\cal L}^{\prime\prime\prime}\left(\frac{2\pi k}{N}\right)\nonumber\\
&=&\frac{1}{2N^{1/2}}\delta_{n_1+n_2+n_3}\sum_{k=1}^{N-1}\prod_{j=1}^3
(e^{2\pi i n_j k/N}-1){\cal L}^{\prime\prime\prime}\left(\frac{2\pi k}{N}\right)\nonumber\\
&\to&\frac{\sqrt{N}}{4\pi}\delta_{n_1+n_2+n_3,0}\int_0^{2\pi} d\theta 
(e^{in_1\theta}-1)(e^{in_2\theta}-1)(e^{in_3\theta}-1){\cal L}^{\prime\prime\prime}(\theta) \ ,
\label{cubic}
\eea
showing that ${\tilde{\cal V}}^3=O(\sqrt{N})$ as desired. A clarification:
Before passing to the continuum approximation the $\delta_{n_1+n_2+n_3}$
is 1 for $n_1+n_2+n_3=0$ mod $N$, with the $n_j$ ranging from $0$ to
$N-1$. We take $N\to\infty$ keeping either
$n_j$ fixed or $N-n_j$ fixed. In the latter case, we replace $n_j$
with the negative integer $n_j-N$, before passing to the continuum
approximation. In the last line the $\delta_{n_1+n_2+n_3,0}$ is 1
for $n_1+n_2+n_3=0$ and mod $N$ is suspended.

The integral on the right side of (\ref{cubic}) may be integrated twice
by parts with vanishing surface terms, using 
\bea
\left[(e^{in_1\theta}-1)(e^{in_2\theta}-1)(e^{in_3\theta}-1)\right]^{\prime\prime}&=&-2i\left[n_1^2\sin n_1\theta+n_2^2\sin n_2\theta+n_3^2\sin n_3\theta\right]
 \ , 
\eea
which requires the constraint $n_1+n_2+n_3=0$. This manipulation shows that
the vertex involves the same integral that occurred in the evaluation
of (\ref{quadratic}):
\bea
I_n\equiv\frac{1}{2\pi}\int_0^{2\pi}\sin n\theta{\cal L}^\prime(\theta)&=&
\begin{cases} 1-(b+f)x^{n}& n\quad {\rm odd}\\
1-(b-f)x^{n}& n\quad {\rm even}\end{cases} \ , \qquad n>0.
\eea
Otherwise  $I_{-n}=-I_n$ and $I_0=0$.

Then the cubic vertex can at $N=\infty$ can be written
\bea
{\tilde{\cal V}}^3_{n_1n_2n_3}&=&-i{\sqrt{N}}\delta_{n_1+n_2+n_3,0}
\left[n_1^2I_{n_1}+n_2^2I_{n_2}+n_3^2I_{n_3}
\right] \ .
\eea
For $x$ near $1/(b+f)$, $n_k$'s equal to $\pm1$ will be important, because their
propagator blows up when $x=1/(b+f)$. For the
cubic vertex, at most two of the $n_k$ can have these values, for example
$n_1=n_2=\pm1$ and $n_3=\mp2$. Since $n_3=0$ is not available, $n_1$ and
$n_2$ must have the same sign.

The generalization to higher $m$ is obvious:
\bea
{\tilde{\cal V}}^m_{n_1\cdots n_m}&\to&\frac{{N}^{2-m/2}}{4\pi}
\delta_{\sum_j n_j,0}\int_0^{2\pi} d\theta 
\prod_{j=1}^m(e^{in_j\theta}-1){\cal L}^{(m)}(\theta) \ .
\eea
Again $m-1$ integrations by parts can be performed with vanishing surface
terms, for which we need
\bea
\frac{d^{m-1}}{d\theta^{m-1}}\prod_{j=1}^m(e^{in_j\theta}-1)=
(-i)^m\sum_{l=1}^{m-1}(-)^l\sum_{k_1<k_2<\cdots<k_l}
(n_{k_1}+\cdots+n_{k_l})^{m-1}\sin(n_{k_1}+\cdots+n_{k_l})\theta \ .
\nonumber
\eea 
Then
\bea
{\tilde{\cal V}}^m_{n_1\cdots n_m}&\to&
-i^m\frac{{N}^{2-m/2}}{2}\delta_{\sum_j n_j,0}
\sum_{l=1}^{m-1}(-)^l\sum_{k_1<k_2<\cdots k_l}
(n_{k_1}+\cdots+n_{k_l})^{m-1}I_{n_{k_1}+\cdots+n_{k_l}} \ .
\eea
\subsection{$1/N$ Perturbations near the Critical Point}
We have defined the $1/N$ expansion in the low temperature phase. But
since there is no actual phase transition at finite $N$, we should be able to
reach the high temperature phase if we keep $N$ finite. The $1/N$
expansion breaks down near $x=1/(b+f)$ because the propagator of
the $n$th mode is $(nN(1-(b+f)x^n))^{-1}$, so the $n=\pm1$ mode
propagator blows up at the critical point. Thus a partial summation
of all orders in the expansion is required. 

First let's examine the lowest order correction to the ``self-energy''
for $n=\pm1$ described by the diagrams in Fig.~\ref{selfen}, keeping
only the lowest contributing mode to each propagator: 
\begin{figure}
\begin{center}\includegraphics[width=4in]{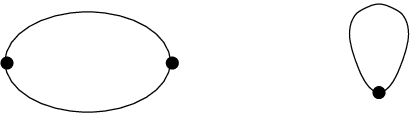}
\caption{Lowest order self energy diagrams.The diagram on the
left represents ${\hat\Pi}^0_1$ and that on the right $\Pi^0_1$.}
\label{selfen}
\end{center}
\end{figure}
\bea
{\hat\Pi}^0_1&=&-\frac{8}{N^2}\frac{(2I_2-I_1)^2}{2I_1I_2}\\
\Pi^0_1&=&\frac{8}{N^2}\frac{2I_2-I_1}{I_1}=\frac{8}{N^2}\frac{1+(b+f)x-2(b-f)x^2}{1-(b+f)x}\\
\Pi^0_1+{\hat\Pi}^0_1&=&\frac{8}{N^2}\frac{2I_2-I_1}{I_1}\left[1-
\frac{2I_2-I_1}{2I_2}\right]=\frac{8}{N^2}\frac{2I_2-I_1}{2I_2} \ ,
\eea
where the normalization is such that the corrected inverse propagator
for the first mode is $N(I_1+{\hat\Pi}^0_1+\Pi^0_1)/2$. We see that both
contributions blow up like $(1-(b+f)x)^{-1}$ at $x=1/(b+f)$. This singularity
in the propagator is responsible for the Hagedorn phenomenon. However at
this order the singularity cancels in the sum of the two contributions!

At finite $N$ a singularity on the real interval $0<x<1$ must be absent, 
which means
the singularity must move off the real axis in the exact propagator. There
is a simple partial summation which accomplishes this which is specified
by a truncated Dyson equation depicted in Fig.~\ref{dyson}:
\begin{figure}
\begin{center}\includegraphics[width=2in]{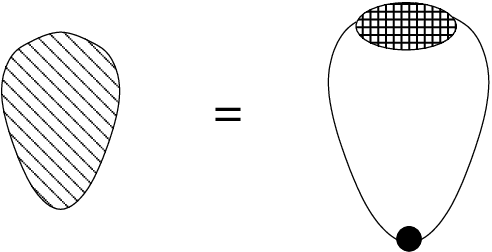}
\caption{Diagrammatic representation of the Dyson equation. The left side 
signifies $\Pi^D$, and the propagator with the shaded bubble represents
the corresponding propagator $\Delta_D=(2/N)(I_1+\Pi^D)^{-1}$.}
\label{dyson}
\end{center}
\end{figure}
\bea
\Pi^D&=&\frac{8}{N^2}\frac{2I_2-I_1}{I_1+\Pi^D} \ .
\eea
This is a quadratic algebraic equation with solutions
\bea
\Pi^D_\pm&=&\frac{1}{2}\left[-I_1\pm\sqrt{I_1^2+\frac{32}{N^2}(2I_2-I_1)}
\right]
\eea
We should choose the plus sign so that $\Pi^D\to0$ as $N\to\infty$,
so the propagator becomes
\bea
\Delta_D&=&\frac{2}{N}\frac{1}{I_1+\Pi^D}=\frac{4}{N}
\left[I_1+\sqrt{I_1^2+\frac{32}{N^2}(2I_2-I_1)}
\right]^{-1} \ .
\label{dysonsol≈ß}
\eea
It is seen that there are no singularities of $\Delta_D$
on the real axis. The original Hagedorn singularity has been replaced
by two complex branch points of order $1/N$ away from the real axis. when
$N\to\infty$, these two branch points coalesce and form the Hagedorn pole.

After this partial summation, the Feynman rules are modified as follows:
(1) The propagator for the first mode is replaced by $\Delta_D$, and
(2) Corrections to the first mode propagator obtained by attaching
any number of the second
diagram of Fig.~\ref{selfen} to any propagator, 
including the propagator in that diagram are deleted. 

At finite $N$
the modified first mode propagator $\Delta_D$ has no singularities on the
real axis in the interval $0<x<1$ corresponding to real temperatures
$0<T<\infty$. The Hagedorn singularity at $T=T_H$ has been replaced
by two branch points off the real axis. These branch points are
responsible for different behavior of the $N\to\infty$ limit depending
on whether $I_1$ is positive or negative. If $I_1>0$ the two terms
inside the square brackets are both positive so the limit gives
\bea
\Delta_D&\sim&\frac{2}{NI_1}\left[1-O\left(\frac{1}{N^2I_1^2}\right)\right],
\qquad I_1>0,\quad N\to\infty \ .
\eea
In contrast if $I_1<0$, the first term in square brackets is negative
and we obtain
\bea
\Delta_D&\sim&\frac{N|I_1|}{4(2I_2-I_1)}+O\left(\frac{1}{N|I_1|}\right),
\qquad I_1<0,\quad N\to\infty.
\eea
We stress that $\Delta_D$ does not include the full second order self
energy since it excludes the first diagram of Fig.\ref{selfen}.
If that diagram had been included in the Dyson equation, the two
branch points would be on the real $x$ axis, one at the zero
of $I_1$, and the other at a slightly higher value of $x$.

\subsection{High Temperature}
Just above the phase transition the $n_j=\pm1$ modes are the only
unstable ones. Then a nonzero vertex, involving only these unstable modes, 
requires that $m$ is even and therefore
half the modes are $+1$ and the other half are $-1$..These 
nonzero vertices reduce to
\bea
{\tilde{\cal V}}^m&=&\frac{{N}^{2-m/2}}{4\pi}
\int_0^{2\pi} d\theta 
(2-2\cos\theta)^{m/2}{\cal L}^{(m)}(\theta) \ .
\eea 

As long as $x<1/(b+f)$ the eigenvalues of $K$ are all negative and of 
order $O(N)$.
In this case the integral over fluctuations limits the $\lambda$'s to
be of order $N^{-1/2}$, so the term $\lambda^m$ is of order $N^{2-m}$.
Thus successive terms in the expansion 
are smaller by a factor of $1/N$. Using this effective field theory,
one can develop
a Feynman graph expansion in which the propagator is determined
by the $\lambda^2$ term and the vertices are ${\tilde{\cal V}}^n$
with $n>2$.

The structure of the perturbation expansion in powers of $1/N$ can also
be used to motivate the large $N$ behavior of the high temperature
phase. When $x=1/(b+f)$, the coefficient of the $\lambda_1^*\lambda_1$
term vanishes and the strength of this mode is not limited by the
quadratic term, so the higher order terms must be relied on to damp the
$\lambda_1$ integral. When $x>1/(b+f)$ the coefficient of the
$\lambda_1^*\lambda_1$ term becomes positive driving toward a higher
maximum. This term becomes comparable to the higher order terms only
for $\lambda_1$ of order $\sqrt{N}$ so the new maximum of $L$ must
be of order $N^2$. Unfortunately, for $\lambda$ this large, all of the
higher order terms are comparable and a perturbative description is
lost. However nonperturbative numerical methods as described in
\cite{sourav} have succeeded in getting a convincing determination of
the high temperature phase.

\section{The Limit $x\to1$: Asymptotics of Eulerian Digraphs}
As noted in \cite{sourav} in the case $b=f=1$ the $x\to1$ limit of the partition function is given by
\begin{equation}
Z(x\rightarrow 1)
=\left(\frac{2}{1-x}\right)^{N-1} \frac{\int_{-\pi}^\pi{\exp\{L(\theta)|_{x=1}\}\prod_{1\leqslant k\leqslant N}{d\theta_k}}}{N!(2\pi)^N}
=\left(\frac{2}{1-x}\right)^{N-1}\frac{R_N}{N!} \ ,
\label{high9}
\end{equation}
where
\bea
R_N=\underset{\mathbb{T}^N}{\text{Res}}\left(\frac{\prod_{1\leqslant i<j\leqslant N}(z_i+z_j)^2}{\prod_{1\leqslant k\leqslant N}(z_k)^N}\right) \ .
\eea
Eq.~\ref{high9} is obtained by substituting
$e^{i\theta_k}\rightarrow z_k$ in the integral. On the other hand,
it is well-known
\cite{McKay1983} that \href{http://oeis.org/A007080}{the number
of labeled Eulerian digraphs with $N$ nodes} is given by $ED(N)$, the coefficient of $
{\textstyle\prod_{j=1}^{N}}
z_{j}^{N-1}$ in
\[
{\textstyle\prod_{j=1}^{N}}
{\textstyle\prod_{k=1+j}^{N}}
\left(  z_{j}+z_{k}\right)^{2} \ .
\]
In other words, in addition to deciding the leading order
$N$-dependence at high temperature, $R_N\equiv ED(N)$ also counts the number
of Eulerian digraphs. 

McKay showed \cite{McKay1990} that
$ED(N)$ is given asymptotically for large $N$ by
\begin{equation}
ED\left(  N\right)  \underset{N\rightarrow\infty}{\sim}\left(  \frac{2^{N}%
}{\sqrt{\pi N}}\right)  ^{N-1}e^{-1/4}\sqrt{N}\left(  1+O\left(  \frac
{1}{\sqrt{N}}\right)  \right)  \ . \label{McKay}%
\end{equation}
We find that the lowest-order correction to this result is%
\begin{equation}
ED\left(  N\right)  \underset{N\rightarrow\infty}{\sim}\left(  \frac{2^{N}%
}{\sqrt{\pi N}}\right)  ^{N-1}e^{-1/4}\sqrt{N}\left(  1+\frac{3}{16N}+O\left(
\frac{1}{N^{2}}\right)  \right)  \ . \label{TLC}%
\end{equation}
We emphasize that the correction is $O\left(  1/N\right)  $ and not $O\left(
1/\sqrt{N}\right)  $. \ In any case, the correction in (\ref{TLC})
significantly reduces the relative error, as shown for $N\leq16$ in the Appendix.

A systematic procedure to obtain the correction in (\ref{TLC}) and higher
order corrections, at least in principle, is described in the following.

\subsection{Methodology}

The basic plan follows that of McKay \cite{McKay1990}\ but
pursues the structure of the ensuing gaussian integrals in more
detail. Applying Cauchy residue theorem to $ED(N)$,  we obtain
an exact integral expression, as given by
\begin{equation}
ED\left(  N\right)  =\left(  \frac{2^{N}}{\pi}\right)  ^{N-1}\int_{-\pi
/2}^{\pi/2}\cdots\int_{-\pi/2}^{\pi/2}~%
{\textstyle\prod_{j=1}^{N-1}}
\left(  \cos^{2}\left(  \alpha_{j}\right)  ~%
{\textstyle\prod_{k=1+j}^{N-1}}
\cos^{2}\left(  \alpha_{j}-\alpha_{k}\right)  \right)  ~d\alpha_{1}%
~\cdots~d\alpha_{N-1}\ . \label{ExactIntegral}%
\end{equation}
Here the $\alpha$ variables are related to the $\theta$'s introduced
earlier by $\alpha_j=\theta_j/2$.
For example, $ED\left(  1\right)  =1$, $ED\left(  2\right)  =2$, $ED\left(
3\right)  =10$, $ED\left(  4\right)  =152$, etc. \ We then write%
\begin{equation}
\cos^{2}\left(  \alpha\right)  =\exp\left(  2\ln\left(  \cos\alpha\right)
\right)  =\exp\left(  -\alpha^{2}-\frac{1}{6}\alpha^{4}-\frac{2}{45}\alpha
^{6}+O\left(  \alpha^{8}\right)  \right)  \label{Truncated}%
\end{equation}
and extend the integrations as $\int_{-\pi/2}^{\pi/2}\rightarrow\int_{-\infty
}^{\infty}$ for any truncated series expansion of $\ln\left(  \cos
\alpha\right)  $\ in the exponential. \ 

A straightforward adaptation of the analysis given by McKay shows, for any
truncation of the series in the exponential (\ref{Truncated}), the additional
contributions coming from $\int_{-\infty}^{-\pi/2}$ and $\int_{\pi/2}^{\infty
}$ will be exponentially suppressed, for large $N$, when compared to the
dominant asymptotic behavior or to any $\left(  1/N\right)  ^{n}$ corrections
to that dominant behavior.

Therefore, to obtain the requisite asymptotic behavior, we need to evaluate
for large $N$%
\begin{equation}
\left\langle \exp\left(  -\frac{1}{6}\sum_{j=1}^{N-1}\alpha_{j}^{4}-\frac
{1}{6}\sum_{j<k}
\left(  \alpha_{j}-\alpha_{k}\right)
^{4}-\frac{2}{45}\sum_{j=1}^{N-1}\alpha_{j}^{6}-\frac{2}{45}\sum_{j<k}%
\left(  \alpha_{j}-\alpha_{k}\right)  ^{6}+O\left(
\alpha^{8}\right)  \right)  \right\rangle \label{EvaluateThis}%
\end{equation}
where the averaging $\left\langle \cdots\right\rangle $\ is to be done with a
\textquotedblleft skewed gaussian\textquotedblright\ measure on $\mathbb{E}%
_{N-1}$, namely,%
\begin{align}
\left\langle f\right\rangle  &  \equiv\int_{-\infty}^{\infty}\cdots
\int_{-\infty}^{\infty}f\left(  \alpha_{1},\cdots,\alpha_{N-1}\right)
~d\mu\ ,\label{Average}\\
d\mu &  =\frac{1}{\sqrt{\pi^{N-1}\left(  \frac{1}{N}\right)  ^{N-2}}}%
~\exp\left(  -\sum_{j=1}^{N-1}\alpha_{j}^{2}-\sum_{j=1}^{N-2}\sum
_{k=1+j}^{N-1}\left(  \alpha_{j}-\alpha_{k}\right)  ^{2}\right)  ~d\alpha
_{1}~d\alpha_{2}~\cdots~d\alpha_{N-1}\ . \label{MeasureAlpha}%
\end{align}
This measure is normalized so that the $\left(  N-1\right)  $-fold integration
$\int_{-\infty}^{\infty}\cdots\int_{-\infty}^{\infty}d\mu=1$ for all $N$. \ 

Now the exponentiated bilinear form in the measure may be diagonalized by
changing variables, as noted by McKay. \ Explicitly, let%
\bea
\alpha_{j}=\frac{1}{\sqrt{N}}~\left(  \beta_{j}+\frac{1}{\left(  1+\sqrt
{N}\right)  }\sum_{k=1}^{N-1}\beta_{k}\right) \ , && \ \ \ \beta_{j}=\sqrt
{N}~\alpha_{j}-\frac{1}{\left(  1+\sqrt{N}\right)  }\sum_{k=1}^{N-1}\alpha
_{k}\ ,\ \ \ \nonumber\\
\alpha_{j}-\alpha_{k}&=&\frac{1}{\sqrt{N}}~\left(  \beta_{j}%
-\beta_{k}\right)  \ .
\eea
This change of variables gives%
\begin{equation}
d\mu=\left(  \frac{1}{\sqrt{\pi}}\right)  ^{N-1}\exp\left(  -\sum_{j=1}%
^{N-1}\beta_{j}^{2}\right)  ~d\beta_{1}~\cdots~d\beta_{N-1}
\label{MeasureGamma}%
\end{equation}
in addition to
\begin{equation}
\sum_{j=1}^{N-2}\sum_{k=1+j}^{N-1}\left(  \alpha_{j}-\alpha_{k}\right)
^{2n}=\frac{1}{N^{n}}\sum_{j=1}^{N-2}\sum_{k=1+j}^{N-1}\left(  \beta_{j}%
-\beta_{k}\right)  ^{2n}%
\end{equation}
for any $n$. \ Indeed, the $\beta$-variables are useful to evaluate the
integrals of the remaining non-diagonal terms in the exponentials, especially
the nested double-summation terms. $\ $But even without this change of
variables it is readily apparent from considering a few examples that
\emph{all positive powers of} $\sum_{j=1}^{N-2}\sum_{k=1+j}^{N-1}\left(
\alpha_{j}-\alpha_{k}\right)  ^{4}$ \emph{must be included in the averages }to
obtain correct asymptotic results to $O\left(  1\right)  $ and to $O\left(
1/N\right)  $, as well as to higher orders in inverse powers of $N$. \ In
particular, as discussed in more detail in the next subsection,
\bea
\left\langle \exp\left(  -\frac{1}{6}\sum_{j=1}^{N-2}\sum_{k=1+j}^{N-1}\left(
\alpha_{j}-\alpha_{k}\right)  ^{4}\right)  \right\rangle
&~&\underset{N\rightarrow\infty}{\sim}~\exp\left(  -\frac{1}{6}\sum_{j=1}%
^{N-2}\sum_{k=1+j}^{N-1}\left\langle \left(  \alpha_{j}-\alpha_{k}\right)
^{4}\right\rangle \right)  +~O\left(  \frac{1}{N}\right)\nonumber\\
&~&\underset{N\rightarrow\infty}{\sim}~e^{-1/4}+~O\left(  \frac{1}{N}\right)
\ . \label{QuarterRoot}%
\eea
This result, along with the normalization factor exhibited in the measure
(\ref{MeasureAlpha}) and the prefactor in (\ref{ExactIntegral}), immediately
gives the dominant asymptotic behavior (\ref{McKay}).

\subsection{First-order Corrections}
Define
\begin{equation}
\left.  \Sigma\right.  _{2n}=\sum_{j=1}^{N-1}\alpha_{j}^{2n}\ ,\ \ \ \left.
\Sigma\Sigma\right.  _{2n}=\sum_{j=1}^{N-2}\sum_{k=1+j}^{N-1}\left(
\alpha_{j}-\alpha_{k}\right)  ^{2n}%
\end{equation}
and find by direct calculation%
\begin{gather}
\left\langle \left.  \Sigma\right.  _{2}\right\rangle =\frac{\left(
N-1\right)  }{N}\ ,\ \ \ \left\langle \left.  \Sigma\right.  _{4}\right\rangle
=\frac{3\left(  N-1\right)  }{N^{2}}\ ,\ \ \ \left\langle \left.
\Sigma\right.  _{6}\right\rangle =\frac{15\left(  N-1\right)  }{N^{3}%
}\ ,\ \ \ \left\langle \left.  \Sigma\right.  _{8}\right\rangle =O\left(
\frac{1}{N^{3}}\right)  \ ,\label{SomeAverages}\\
\left\langle \left.  \Sigma\Sigma\right.  _{2}\right\rangle =\frac{\left(
N-1\right)  \left(  N-2\right)  }{2N}\ ,\ \ \ \left\langle \left.
\Sigma\Sigma\right.  _{4}\right\rangle =\frac{3\left(  N-1\right)  \left(
N-2\right)  }{2N^{2}}\ ,\ \ \ \nonumber\\
\left\langle \left.  \Sigma\Sigma\right.
_{6}\right\rangle =\frac{15\left(  N-1\right)  \left(  N-2\right)  }{2N^{3}%
}\ ,\ \ \ \left\langle \left.  \Sigma\Sigma\right.  _{8}\right\rangle
=O\left(  \frac{1}{N^{2}}\right)  \ .\nonumber
\end{gather}
Moreover, again by direct calculation,%
\begin{gather}
\left\langle \left(  \left.  \Sigma\Sigma\right.  _{4}\right)  ^{2}%
\right\rangle =\frac{\left(  N-1\right)  \left(  N-2\right)  }{N^{4}}\left(
\frac{9}{4}N^{2}+\frac{51}{4}N-6\right)  \ ,\label{SigmaSigma4Powers}\\
\left\langle \left(  \left.  \Sigma\Sigma\right.  _{4}\right)  ^{3}%
\right\rangle =\frac{\left(  N-1\right)  \left(  N-2\right)  }{N^{6}}\left(
\frac{27}{8}N^{4}+\frac{135}{2}N^{3}+\frac{3267}{8}N^{2}-\frac{837}%
{4}N+54\right)  \ ,\nonumber\\
\left\langle \left(  \left.  \Sigma\Sigma\right.  _{4}\right)  ^{4}%
\right\rangle =\frac{\left(  N-1\right)  \left(  N-2\right)  }{N^{8}}\left(
\frac{81}{16}N^{6}+\frac{3483}{16}N^{5}+\frac{58\,185}{16}N^{4}+\frac
{413\,505}{16}N^{3}-\frac{53\,703}{8}N^{2}\right.\nonumber\\
\left.+\frac{18\,819}{2}N-3240\right)
\ ,\nonumber
\end{gather}
etc. \ Also, while it is not required for the problem at hand, it is not too
difficult to show for any integer $n$ that%
\begin{equation}
\left\langle \left(  \left.  \Sigma\Sigma\right.  _{2}\right)  ^{n}%
\right\rangle =\frac{\Gamma\left(  \frac{1}{2}N-1+n\right)  }{\Gamma\left(
\frac{1}{2}N-1\right)  }~\frac{\left(  N-1\right)  ^{n}}{N^{n}}\ .
\label{SigmaSigma2Powers}%
\end{equation}
The asymptotic behavior of this last result nicely illustrates some features
that are relevant to the problem at hand. \ An averaged exponential of
$\left.  \Sigma\Sigma\right.  _{2}$ gives%
\begin{equation}
\left\langle \exp\left(  -\lambda\left.  \Sigma\Sigma\right.  _{2}\right)
\right\rangle =\left(  1+\frac{\lambda\left(  N-1\right)  }{N}\right)
^{1-\frac{1}{2}N}%
\end{equation}
while an averaged exponential of $\left(  \left.  \Sigma\Sigma\right.
_{2}\right)  ^{2}$ gives%
\begin{equation}
\left\langle \exp\left(  -\lambda\left(  \frac{\left.  \Sigma\Sigma\right.
_{2}}{N}\right)  ^{2}\right)  \right\rangle =\sum_{n=0}^{\infty}\frac
{\Gamma\left(  \frac{1}{2}N-1+2n\right)  }{\Gamma\left(  \frac{1}%
{2}N-1\right)  }~\frac{\left(  1-N\right)  ^{2n}}{N^{2n}}\frac{(-\lambda/N^2)^{n}%
}{n!}\ .
\end{equation}
The latter series is \emph{divergent}. \ This is not unexpected
\footnote{For example, consider just a single integral as follows.%
\[
I\left(  a,b\right)  =\int_{0}^{\infty}e^{-ax^{2}-bx^{4}}dx=\frac{1}{2\sqrt
{a}}\sum_{n=0}^{\infty}\frac{\left(  -\frac{b}{a^{2}}\right)  ^{n}}{n!}%
~\Gamma\left(  2n+\frac{1}{2}\right)  =\frac{1}{2}\sqrt{\frac{\pi}{a}%
}\operatorname{hypergeom}\left(  \left[  \frac{1}{4},\frac{3}{4}\right]
,\left[  {}\right]  ,-4\frac{b}{a^{2}}\right)  \ .
\]
This is a divergent series in powers of $b$. \ However, as an expansion in
$a$, a convergent series is obtained.%
\[
I\left(  a,b\right)  =\frac{1}{4b^{1/4}}\sum_{n=0}^{\infty}\frac{\left(
-\frac{a}{\sqrt{b}}\right)  ^{n}}{n!}~\Gamma\left(  \frac{n+1/2}{2}\right)
=\frac{\pi}{8}\sqrt{\frac{2a}{b}}~e^{\frac{a^{2}}{8b}}\left(  I_{-1/4}\left(
\frac{a^{2}}{8b}\right)  -I_{1/4}\left(  \frac{a^{2}}{8b}\right)  \right)
\ .
\]
The first of these series correctly gives the asymptotic expansion of
$I\left(  a,b\right)  $ for large $a$ with $b$ fixed, while the second nicely
exhibits the large $b$ behavior for any fixed $a$.
} \ Nevertheless, asymptotically for large $N$,%
\begin{equation}
\left\langle \left(  \frac{\left.  \Sigma\Sigma\right.  _{2}}{N}\right)  ^{2n}%
\right\rangle =\frac{1}{4^n}\left(  1+\frac{4n\left(  n-2\right)  }{N}%
+\frac{n\left(  24n^{3}-112n^{2}+132n-29\right)  }{3N^{2}}+O\left(  \frac
{1}{N^{3}}\right)  \right)
\end{equation}
and each of the individual terms in the asymptotic behavior can be
exponentially summed as convergent series. \ For example,%
\begin{align}
\sum_{n=0}^{\infty}\frac{1}{n!}\left(  \frac{-\lambda}{4}\right)  ^{n}  &
=e^{-\frac{1}{4}\lambda}\ ,\\
\frac{4}{N}\sum_{n=0}^{\infty}\frac{1}{n!}\left(  \frac{-\lambda}{4}\right)
^{n}n\left(  n-2\right)   &  =e^{-\frac{1}{4}\lambda}\left(  4+\lambda\right)
\left(  \frac{\lambda}{4N}\right)  \ ,\\
\frac{1}{3N^{2}}\sum_{n=0}^{\infty}\frac{1}{n!}\left(  \frac{-\lambda}%
{4}\right)  ^{n}n\left(  24n^{3}-112n^{2}+132n-29\right)   &  =\nonumber\\
& \hskip-1ine^{-\frac{1}%
{4}\lambda}\left(  -15-9\lambda-2\lambda^{2}+\frac{3}{8}\lambda^{3}\right)
\left(  \frac{\lambda}{12N^{2}}\right)  \ .
\end{align}
Thus we obtain as a warm-up exercise
\bea
\left\langle \exp\left(  -\lambda\left(  \frac{\left.  \Sigma\Sigma\right.
_{2}}{N}\right)  ^{2}\right)  \right\rangle &\underset{N\rightarrow\infty}{\sim
}&\nonumber\\
&&\hskip-1.5ine^{-\frac{1}{4}\lambda}\left(  1+\frac{\lambda}{4N}\left(  4+\lambda\right)
+\frac{\lambda}{12N^{2}}\left(  -15-9\lambda-2\lambda^{2}+\frac{3}{8}%
\lambda^{3}\right)  +O\left(  \frac{1}{N^{3}}\right)  \right)  \ .
\eea
Similar techniques can now be used to obtain (\ref{TLC}), but for that result
we need to evaluate $\left\langle \left(  \left.  \Sigma\Sigma\right.
_{4}\right)  ^{n}\right\rangle $. \ This is not as simple as
(\ref{SigmaSigma2Powers}).

In general, as is evident from the examples in (\ref{SigmaSigma4Powers}),
\begin{equation}
\left\langle \left(  \left.  \Sigma\Sigma\right.  _{4}\right)  ^{n}%
\right\rangle =\frac{\left(  N-1\right)  \left(  N-2\right)  }{N^{2n}}%
\sum_{k=0}^{2\left(  n-1\right)  }A_{k}\left(  n\right)  ~N^{k}\ .
\end{equation}
All the $A_{k}\left(  n\right)  $ coefficients in this case are not yet
available. \ Fortunately, however, to obtain (\ref{TLC}) only the two leading
coefficients are required, namely, $A_{2n-2}$ and $A_{2n-3}$. \ These are
\emph{not} so difficult to obtain. \ By direct calculation the leading large
$N$ coefficient is $A_{2n-2}=\left(  \frac{3}{2}\right)  ^{n}$, followed by
$A_{2n-3}=\left(  \frac{3}{2}\right)  ^{n}\frac{1}{3}\left(  n-1\right)
\left(  13n-9\right)  $. 
 Therefore%
\begin{equation}
\left\langle \left(  \left.  \Sigma\Sigma\right.  _{4}\right)  ^{n}%
\right\rangle \underset{N\rightarrow\infty}{\sim}\left(  \frac{3}{2}\right)
^{n}\left(  1+\frac{13n^{2}-22n}{3N}+O\left(  \frac{1}{N^{2}}\right)  \right)
\ ,
\end{equation}
and these leading terms may be exponentially summed to obtain%
\begin{equation}
\left\langle \exp\left(  -\lambda\left.  \Sigma\Sigma\right.  _{4}\right)
\right\rangle \underset{N\rightarrow\infty}{\sim}e^{-\frac{3}{2}\lambda
}\left(  1+\frac{18\lambda+39\lambda^{2}}{4N}+O\left(  \frac{1}{N^{2}}\right)
\right)  \ .
\end{equation}
In particular, for $\lambda=1/6$,%
\begin{equation}
\left\langle \exp\left(  -\frac{1}{6}\left.  \Sigma\Sigma\right.  _{4}\right)
\right\rangle \underset{N\rightarrow\infty}{\sim}e^{-\frac{1}{4}}\left(
1+\frac{49}{48N}+O\left(  \frac{1}{N^{2}}\right)  \right)  \ .
\label{SigmaSigma4}%
\end{equation}

In view of the above results, the asymptotic behavior of interest is%
\begin{align}
&  \left\langle \exp\left(  -\frac{1}{6}\left.  \Sigma\right.  _{4}-\frac
{1}{6}\left.  \Sigma\Sigma\right.  _{4}-\frac{2}{45}\left.  \Sigma\right.
_{6}-\frac{2}{45}\left.  \Sigma\Sigma\right.  _{6}+O\left(  \alpha^{8}\right)
\right)  \right\rangle \\
&  \underset{N\rightarrow\infty}{\sim}\left\langle \left(  1-\frac{1}%
{6}\left.  \Sigma\right.  _{4}-\frac{2}{45}\left.  \Sigma\Sigma\right.
_{6}+O\left(  \frac{1}{N^{2}}\right)  \right)  \exp\left(  -\frac{1}{6}\left.
\Sigma\Sigma\right.  _{4}\right)  \right\rangle \ .
\end{align}
The \textquotedblleft$1$\textquotedblright\ term is given by
(\ref{SigmaSigma4}) up to but not including $O\left(  1/N^{2}\right)  $. \ But
then two more averages involving $\exp\left(  -\left.  \Sigma\Sigma\right.
_{4}/6\right)  $ are needed. \ They are%
\begin{align}
&  \left\langle \left(  \left.  \Sigma\right.  _{4}\right)  \exp\left(
-\frac{1}{6}\left.  \Sigma\Sigma\right.  _{4}\right)  \right\rangle
\underset{N\rightarrow\infty}{\sim}3~e^{-\frac{1}{4}}\left(  \frac{1}%
{N}+O\left(  \frac{1}{N^{2}}\right)  \right)  \ ,\\
&  \left\langle \left(  \left.  \Sigma\Sigma\right.  _{6}\right)  \exp\left(
-\frac{1}{6}\left.  \Sigma\Sigma\right.  _{4}\right)  \right\rangle
\underset{N\rightarrow\infty}{\sim}\frac{15}{2}~e^{-\frac{1}{4}}\left(
\frac{1}{N}+O\left(  \frac{1}{N^{2}}\right)  \right)  \ .
\end{align}
Finally, $49/48-1/2-1/3=3/16$ and we obtain%
\bea
\left\langle \exp\left(  -\frac{1}{6}\left.  \Sigma\right.  _{4}-\frac{1}%
{6}\left.  \Sigma\Sigma\right.  _{4}-\frac{2}{45}\left.  \Sigma\right.
_{6}-\frac{2}{45}\left.  \Sigma\Sigma\right.  _{6}+O\left(  \alpha^{8}\right)
\right)  \right\rangle &\underset{N\rightarrow\infty}{\sim}&\nonumber\\
&&\hskip-1in e^{-\frac{1}{4}%
}\left(  1+\frac{3}{16N}+O\left(  \frac{1}{N^{2}}\right)  \right)
\label{3sixteenths}%
\eea
These results complete the derivation of the first-order correction. \ Putting
everything together --- the normalization factor exhibited in the measure
(\ref{MeasureAlpha}), the prefactor in (\ref{ExactIntegral}), and
(\ref{3sixteenths}) --- we obtain (\ref{TLC}).

\subsection{Central Moments}

In our opinion, it would not be surprising if the central limit theorem
underlies some of the results given above. \ Perhaps the following comments
shed some light on this issue.

Define the $n$th central moment of $f$ as%
\begin{equation}
\sigma_{n}\left[  f\right]  \equiv\left\langle \left(  f-\left\langle
f\right\rangle \right)  ^{n}\right\rangle =\sum_{k=0}^{n}\binom{n}{k}\left(
-1\right)  ^{k}\left\langle f^{n-k}\right\rangle \left\langle f\right\rangle
^{k}%
\end{equation}
where $\left\langle f^{0}\right\rangle \equiv1$, and consider central moments
of $\left.  \Sigma\Sigma\right.  _{4}$. \ For example, from
(\ref{SomeAverages}) and (\ref{SigmaSigma4Powers}),%
\begin{equation}
\sigma_{2}\left[  \left.  \Sigma\Sigma\right.  _{4}\right]  =\frac{3}{2N^{4}%
}\left(  N-1\right)  \left(  N-2\right)  \left(  13N-7\right)  =\frac{39}%
{2N}+O\left(  \frac{1}{N^{2}}\right)
\end{equation}
which is $O\left(  1/N\right)  $ compared to $O\left(  1\right)  $ for
$\left\langle \left(  \left.  \Sigma\Sigma\right.  _{4}\right)  ^{2}%
\right\rangle $. \ On the other hand, for all $n\geq3$, We find%
\begin{equation}
\sigma_{n}\left[  \left.  \Sigma\Sigma\right.  _{4}\right]  =O\left(  \frac
{1}{N^{2}}\right)  \label{CentralMomentBeyond2}%
\end{equation}
as opposed to $\left\langle \left(  \left.  \Sigma\Sigma\right.  _{4}\right)
^{n}\right\rangle =O\left(  1\right)  $. \ This fact alone permits a recursive
calculation of both the $O\left(  1\right)  $ and the $O\left(  1/N\right)  $
terms in $\left\langle \left(  \left.  \Sigma\Sigma\right.  _{4}\right)
^{n}\right\rangle $ for $n\geq3$.

It could be interesting to exploit additional relations of this sort to
determine higher-order corrections to the asymptotic behavior of $ED\left(
N\right)  $. \ While research on the number of digraphs has continued up to
the present, as evident in \cite{Recent} and references cited therein,
higher-order asymptotic corrections are as yet unexplored.

\section{Conclusion}
We have gone some distance toward understanding the physics of string
bits at finite temperature. In particular we have elucidated the concept
of a Hagedorn transition in a system whose only dynamics is a 
restriction to the singlet sector. One can think of this simplified
dynamics as underlying a tensionless string, since the $P^-$ of
the emergent string, proportional to $T_0$ is zero. Extending the
analysis to a string bit model with $T_0>0$ is a clear direction for
future research.

The partition function of the system studied here also has the
purely group theoretic interpretation as a generating function 
for the number
of color singlet states in a system of bosonic and fermionic creation operators
each in the adjoint representation. For finite $N$ the partition function
is finite and smooth in the whole range $0<x<1$. But only in what we have 
called the low temperature phase $x<1/(b+f)$ does the finite $N$ partition
function have a finite $N\to\infty$ limit. The corresponding limit for
$x>1/(b+f)$ does not exist, but its logarithm grows quadratically with
$N$. In particular in the case $b=f=1$ at $x=1$ the partition function is
related to known counting functions for digraphs, which
have been successfully analyzed for large $N$ \cite{McKay1983,McKay1990,sourav},
and for which we have calculated $1/N$ corrections in this article..

We have briefly touched on setting up a systematic $1/N$ expansion for our
system. For the low temperature phase it is well defined and analytically
tractable and deserves further study. It would also be of interest to
do a better job on the high temperature phase for $x_H<x<1$. In the 
context of effective field theory about the uniform $\theta$ 
distribution characterizing the low temperature phase, some 
qualitative information
about the system near the critical point can be gleaned, but we
have not been able to draw rigorous quantitative conclusions.
However the numerical studies of \cite{sourav} and Section 6 of
the present article give a rather
detailed account of the physics of the high temperature phase. 
\medskip

\noindent\textbf{Acknowledgements}
We thank David McGady for drawing our attention to the
references \cite{sundborg,aharony}
and for valuable comments about the issues we have discussed.
We also thank Brendan McKay for his comments and for
pointing out more recent literature. \ This work was supported in part by a
University of Miami Cooper Fellowship, in part by a Clark Way Harrison Visiting
Professorship, and in part by the Department of Energy under Grant No. DE-SC0010296.


\bigskip
\subsection*{Appendix: \ Relative Errors}

The relative error is given by
\begin{equation}
\Delta\left(  N\right)  =\left(  ED\left(  N\right)  -ED_{\text{asymp}}\left(
N\right)  \right)  /ED\left(  N\right)  \nonumber\label{A1}%
\end{equation}
where $ED\left(  N\right)  $ is exact and $ED_{\text{asymp}}\left(  N\right)
$\ is either McKay's result (\ref{McKay}) or the corrected result (\ref{TLC}).
\ Numerical results for $1\leq N\leq16$ are given in the following Table. \ To
compute these relative errors, exact results for $ED\left(  N\right)  $ for
$N\leq16$ were taken from \cite{McKay1983}.

\begin{center}
$%
\begin{array}
[c]{ccc}%
\fbox{$N$} & \fbox{$\Delta\left(  N\right)  $ using $ED_\text{asymp}$ in
(\ref{McKay})} & \fbox{$\Delta\left(  N\right)  $ using $ED_\text{asymp}$ in
(\ref{TLC})}\\
&  & \\
1 & 22.\,\allowbreak120\times10^{-2} & 7.\,\allowbreak517\times10^{-2}\\
2 & 12.\,\allowbreak122\times10^{-2} & 3.\,\allowbreak883\times10^{-2}\\
3 & 8.\,\allowbreak400\times10^{-2} & 2.\,\allowbreak675\times10^{-2}\\
4 & 5.\,\allowbreak777\times10^{-2} & 1.\,\allowbreak360\times10^{-2}\\
5 & 4.\,\allowbreak335\times10^{-2} & 7.\,\allowbreak471\times10^{-3}\\
6 & 3.\,\allowbreak496\times10^{-2} & 4.\,\allowbreak801\times10^{-3}\\
7 & 2.\,\allowbreak939\times10^{-2} & 3.\,\allowbreak390\times10^{-3}\\
8 & 2.\,\allowbreak537\times10^{-2} & 2.\,\allowbreak531\times10^{-3}\\
9 & 2.\,\allowbreak233\times10^{-2} & 1.\,\allowbreak965\times10^{-3}\\
10 & 1.\,\allowbreak995\times10^{-2} & 1.\,\allowbreak571\times10^{-3}\\
11 & 1.\,\allowbreak802\times10^{-2} & 1.\,\allowbreak285\times10^{-3}\\
12 & 1.\,\allowbreak644\times10^{-2} & 1.\,\allowbreak071\times10^{-3}\\
13 & 1.\,\allowbreak511\times10^{-2} & 9.\,\allowbreak064\times10^{-4}\\
14 & 1.\,\allowbreak398\times10^{-2} & 7.\,\allowbreak771\times10^{-4}\\
15 & 1.\,\allowbreak301\times10^{-2} & 6.\,\allowbreak737\times10^{-4}\\
16 & 1.\,\allowbreak217\times10^{-2} & 5.\,\allowbreak896\times10^{-4}%
\end{array}
$
\end{center}

\end{document}